# Lightweight Security for Network Coding

João P. Vilela    Luísa Lima    João Barros

Instituto de Telecomunicações
Department of Computer Science
Universidade do Porto, Porto, Portugal
{joaovilela, luisalima, barros}@dcc.fc.up.pt

*Abstract*—Under the emerging network coding paradigm, intermediate nodes in the network are allowed not only to store and forward packets but also to process and mix different data flows. We propose a low-complexity cryptographic scheme that exploits the inherent security provided by random linear network coding and offers the advantage of reduced overhead in comparison to traditional end-to-end encryption of the entire data. Confidentiality is achieved by protecting (or "locking") the source coefficients required to decode the encoded data, without preventing intermediate nodes from running their standard network coding operations. Our scheme can be easily combined with existing techniques that counter active attacks.

## I. INTRODUCTION

Since the seminal paper of Ahlswede, Li, Cai and Yeung [1], where it is proved that the max-flow min-cut capacity of a general multicast network can only be achieved by allowing intermediate nodes to mix different data flows, a surge in *network coding* research (e.g. [2], [3], [4]) has uncovered its potential to provide higher throughput and robustness.

The basic idea behind network coding is illustrated in *Figure 1*. Suppose that node 1 aims at sending bits $a$ and $b$ simultaneously (i.e. multicast) to sinks 6 and 7. It is not difficult to see that the link between nodes 4 and 5 results in a bottleneck in the sense that either bit $a$ is forwarded (in which case node 6 does not receive bit $b$), or bit $b$ is sent (in which case node 7 will receive incomplete information). It follows that although the capacity of the network is 2 bits per transmission (because the min-cut to each destination equals 2), this cannot be achieved unless node 4 jointly encodes $a$ and $b$, for example, through an XOR operation that allows perfect recovery at the sinks.

It turns out that random linear network coding (RLNC), explained in detail in the next section, is sufficient to reach the multicast capacity of a network [5], [2], [4]. In addition, robustness gains have been reported in packetized networks with lossy links in [3]. The transition from theory to practice is well illustrated by the Practical Network Coding scheme proposed in [6]. Recent successful applications include peer-to-peer networks (e.g. for content distribution [7], [8]) and wireless networks [9], [10].

It is fair to say that current proposals focusing on security aspects of network coding are mainly of theoretical nature. References [11] and [12] present a secure linear network code that achieves perfect secrecy against an external eavesdropper

The first two authors contributed equally to this work, which was partly supported by the Fundação para a Ciência e Tecnologia (Portuguese Foundation for Science and Technology) under grants SFRH/BD/28056/2006 and SFRH/BD/24718/2005.

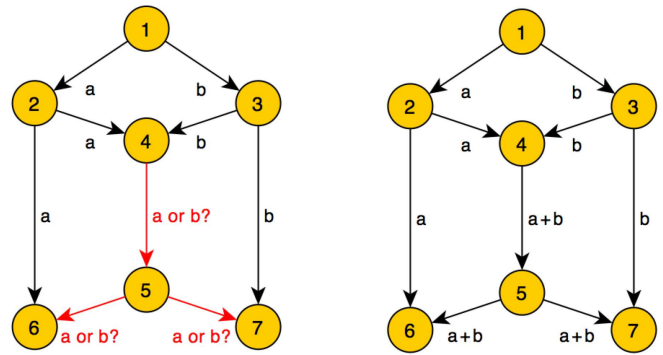

Fig. 1. Canonical network coding example: node 1 multicasts bits $a$ and $b$ to nodes 6 and 7. If node 4 did not perform a simple encoding operation on the incoming bits, the max-flow min-cut capacity of 2 could not be achieved.

with access to a limited number of links. A different threat posed by "nice but curious" intermediate nodes is analyzed in [13]. Active attacks, in particular Byzantine modifications, are considered in [14], [15] and [16].

Motivated by the security challenges of emerging network coded systems we present a lightweight cryptographic scheme to ensure confidentiality in network coding, which leverages the inherent security provided by RLNC to reduce the overhead in comparison to end-to-end encryption of the entire data flow. The main novelty of our approach lies in protecting (or "locking") the source coefficients required to decode the linearly coded data, while still allowing intermediate nodes to carry out the usual network coding operations on a set of coefficients containing the "unlocked" coefficients followed by the "locked" coefficients. Thus, our security mechanisms can be combined with state-of-the-art network coding protocols without the need to modify the coding procedures at the intermediate nodes. As part of our contribution, we discuss in detail the threats countered and the computational overhead incurred by our secure network coding scheme.

The remainder of the paper is organized as follows. First, Section II describes the network model, the RLNC paradigm, and a practical RLNC protocol. Then, Section III explains our security scheme in detail. Its performance is evaluated in Section IV and the paper concludes with Section V.

## II. MODELS AND ASSUMPTIONS

We start by describing our network assumptions, the basics of RLNC and the threat model based upon which our security design is built.

## A. Network Assumptions

We consider a network abstraction where the source and intermediate nodes have access to the identifiers of the sinks (e.g. the IP addresses). This way, our schemes can be easily adapted to the many classes of networks that share this characteristic, in particular networks with no centralized knowledge of the network topology or of the encoding functions. It is worth pointing out that network coding has been proposed at several different layers of the protocol stack, for instance [6] addresses the network layer, whereas [7], [8] focus on the application layer. Cross-layer protocols appear in [10].

We further assume that the source and sink nodes share symmetric keys to encrypt data as needed. Several mechanisms can be used for the exchange of shared keys, such as an offline mechanism for pre-distribution of keys, an authentication protocol such as Kerberos or a Public Key Infrastructure (PKI).

## B. Random Linear Network Coding

Random Linear Network Coding (RLNC) [4] is a distributed methodology for performing network coding, in which each node in the network independently and randomly selects a set of coefficients and uses them to form linear combinations of the data packets it receives. As shown in *Figure 2*, these linear combinations are then sent over the outgoing links. In the example, $\alpha_i$ and $\beta_j$ are chosen at random from a finite field. Each packet is sent along with the global encoding vector [2], which is the set of linear transformations that the original packet goes through on its path from the source to the destination. The global encoding vector enables the receivers to decode the original data using Gaussian elimination. If the coefficients are chosen at random from a large enough field, the resulting matrix is invertible with high probability, which explains why this approach is capable of achieving the multicast capacity of a network.

A framework for packetized network coding (*Practical Network Coding*, PNC) is presented in [6], which leverages RLNC's resilience against disruptions such as packet loss, congestion, and changes of topology in order to guarantee robust communication over highly dynamic networks with minimal (or no) control information. The most important items in the framework are the packet format and the buffering model.

The packet format consists of the *global encoding vector* (kept in the header) and the payload, which is divided into vectors according to the field size ($2^8$ or $2^{16}$, i.e. each symbol has 8 or 16 bits, respectively). Each of these symbols is then used as a building block for the linear operations performed by the nodes.

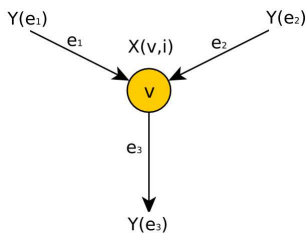

Fig. 2. Linear operations at intermediate node $v$ in the network. $X(v, i)$ represents the traffic generated at node $v$, $Y(e_1)$ and $Y(e_2)$ represent traffic in the incoming links $e_1$ and $e_2$, respectively, and $Y(e_3)$ represent traffic in the outgoing link $e_3$. In linear network coding, we have that $Y(e_3) = \sum_i \alpha_i X(v,i) + \sum_{j=1,2} \beta_j Y(e_j)$ [2].

The buffering model divides the stream of packets into *generations* of size $h$, such that packets in the same generation are tagged with a common generation number. Each node sorts the incoming packets in a single buffer according to their generation number. When there is a transmission opportunity at an outgoing edge, the sending node generates a new packet, which contains a random linear combination of all packets in the buffer that belong to the *current* generation. If a packet is *non-innovative*, i.e. if it does not increase the rank of the decoding matrix available at the receiving node, then it is immediately discarded. As soon as the matrix of received packets has full rank, Gaussian elimination is performed at the receivers to obtain the original packets.

## C. Threat Model

We consider the threat posed by an attacker with the following characteristics:

1) he can observe every transmission in the network;
2) he has full access to information about the encoding and decoding schemes;
3) he is computationally bounded and thus unable to break hard cryptographic primitives;
4) he can drop or erase packets in the network at will (traffic relay refusal);
5) he can inject traffic in the network at will which allows him, for example, to introduce bogus packets in the network to decrease the diversity and robustness of the system.

The focus of our contribution is on threats (1) – (3). Characteristics (4) and (5) are dealt with at the end of Section III.

## III. SECURITY SCHEME

We propose SPOC (Secure Practical Network Coding), a security framework that exploits the interplay between the intrinsic security of network coding and standard cryptographic mechanisms with the goal of countering the threats described in *Section II*.

To deal with characteristics 1–3 of the attacker profile, SPOC introduces modifications to RLNC based protocols (e.g. PNC) only at the source and receiver nodes. We define two types of coefficients: (1) the *unlocked coefficients*, which are basically a line of coefficients drawn from the identity matrix for each coded packet, and (2) the *locked coefficients*, which are actually used for encoding and decoding yet are encrypted with keys that are available at the destination. The unlocked and locked coefficients are concatenated and added to the packet header whenever a new packet is generated.

The full set of coefficients (locked and unlocked) is processed by the intermediate nodes following the exact same packet mixing rules of the original RLNC based protocol. In other words, there is no need to change the protocol at the intermediate nodes. This is made possible by using an encryption system whose output ciphertext is of the same size as the plaintext, e.g. AES. Our approach thus has the advantage that encryption of coefficients is required only once for each packet in a certain generation (see *Section II*).

At the destination nodes, the unlocked coefficients are of major relevance since they store the operations performed along the network upon the locked coefficients. Only after reversing the operations performed along the network, can the destination nodes decrypt the locked coefficients and, thus, have access to the hidden information.

It is important to stress that the unlocked coefficients do not provide any information for effectively decoding the packets without access to a decrypted version of the locked coefficients – they only indicate whether the packets are linearly independent or not, and are used for the execution of several steps of original RLNC based protocols. Moreover, the payload can be deemed to be secure against the described threats since each symbol results from a linear combination with random coefficients that are locked by our scheme and thus inaccessible to the attacker.

Apart from the typical requirements for encryption mechanisms, such as efficiency and strong security, our scheme requires an encryption mechanism with ciphertext size equal to that of the plaintext. This enables intermediate nodes to run operations on the locked coefficients without the need for the decryption keys. To exemplify this, let us consider symbols with 8 bits (i.e. field size of $2^8$). For each packet sent in a certain generation, the $h$ coefficients are concatenated resulting in a string of $h$ bytes of plaintext. The string of plaintext coefficients is then encrypted (or "locked") as a whole, resulting in $h$ bytes of ciphertext. The ciphertext is then divided into $h$ blocks of 8 bits, each of them corresponding to an encrypted locked coefficient ready to be sent in a packet.

To illustrate the basic principles of the proposed scheme, *Figure 3* presents the canonical network coding example of *Section I* with the modifications introduced by our approach. The operations in this example can be described as follows.

1) The *source node 1* randomly generates the locked coefficients and encrypts them with the keys shared with the sink nodes *6* and *7* using one of the already mentioned cryptosystems. The unlocked coefficients of each packet are simply distinct lines of the identity matrix which allow subsequent nodes to check if the packets are linearly independent or not, and carry out the protocol using this information;
2) The subsequent *intermediate nodes* perform the usual combination of packets (e.g. node *4* combines the packets received from nodes *2* and *3* using the $(1, 1)$ encoding vector). The intermediate nodes do not differentiate between the locked and unlocked coefficients, in that they perform exactly the same operations on both;
3) When sufficient information reaches the *destination nodes 6* and *7*, they recover the original locked coefficients using the knowledge of the transformation they suffered – available through the unlocked coefficients. Then, they decrypt the locked coefficients and compute the product with the unlocked coefficients. The destination nodes finally perform Gaussian elimination to recover the native packets.

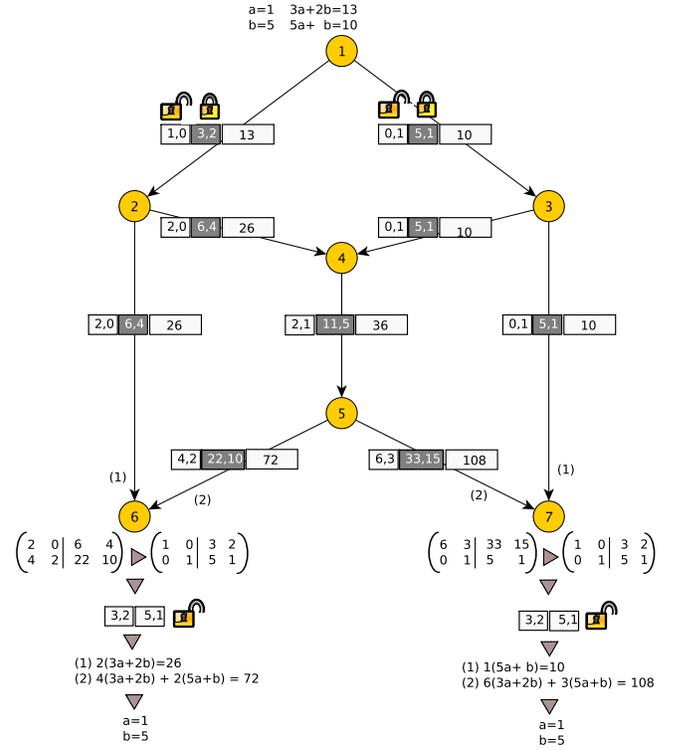

Fig. 3. Basic scheme. White parts of packets represent clear-text information, whereas encrypted information is shaded in grey. In practice, the initial locked coefficients in grey – (3,2) and (5,1) – are encrypted to other symbols of the same size. For illustration purposes, the scheme is simplified using integers.

A formal description of our scheme is provided in *Table I*.

As mentioned above, we consider two types of active attacks ((4) and (5) in the list of *Section II*). The first type of attack, *traffic relay refusal*, can be viewed as a special case of loss or erasure of packets in a network. Their impact is already reduced by the properties of RLNC (see [3]), from which our protocol benefits. As for the second type of attack, i.e. the injection of malicious bogus traffic in the network, we can extend the "Shared Secret Model" from the protocol presented in [14] in the following way. The source and sinks share secret keys, which can be used to share some extra redundancy on the original information sent by the source. This helps the sinks to infer, for each generation, the modifications introduced by the active attacker on the packets flowing in the network. For each generation, the source sends a secret composed of a parity-check matrix and a hash matrix equal to its product with the matrix of information generated by the source. The information present in the unlocked coefficients and the shared secret is then used at the sinks to detect active attacks and decode the original information. This solution achieves a rate of $h - z_0$, where $h$ is the original capacity of the network – the number of packets in a generation – and $z_0$ is the rate of packets injected by the adversary.

TABLE I
SUMMARY OF SPOC

**Initialization (source nodes):**
- A key management mechanism is used to exchange shared keys with the sink nodes, which are used for the encryption of the locked coefficients (see also *Section II*).
- The source node stores the message packets $w_1, w_2, ..., w_h$ in its memory;
- The source node forms a random linear combination of the $h$ packets in its memory (the current generation) and puts it in a packet to be sent;
- The coefficients corresponding to a distinct line of the $h \times h$ identity matrix are added to the header of each coded packet. These correspond to the *unlocked* coefficients;
- The packet's global encoding vector is encrypted with the shared keys and also placed in the header of each packet. These correspond to the *locked* coefficients.

**Operation at intermediate nodes:**
- When a packet is received by a node, the node stores the packet in its memory;
- To transmit on an outgoing link, the node produces a packet by forming a random linear combination of the packets in its buffer, modifying both the unlocked and locked coefficients without distinction, according to the rules of standard RLNC based protocols.

**Decoding (sink nodes):**
- When *sufficient packets are received*:
  – Using the unlocked coefficients (which store the operations performed upon the locked coefficients throughout the network), the receiver reverts those operations thus obtaining the original locked coefficients;
  – The receiver then decrypts the locked coefficients;
  – The receiver determines the decoding matrix by computing the product of the unlocked coefficients and the corresponding locked coefficients;
  – Gaussian elimination is then performed to recover the original packets.

TABLE II
VOLUME OVERHEAD OF LOCKED COEFFICIENTS (PER PACKET).

| MAXIMUM IP PACKET SIZE | #CODED PACKETS $h$ | OVERHEAD IN $\mathbb{F}_q$ $q=2^8$ | $q=2^{16}$ |
|---|---|---|---|
| 1500 | 20 | 1.3% | 2.7% |
|  | 50 | 3.3% | 6.7% |
|  | 100 | 6.7% | 13.3% |
|  | 200 | 13.3% | 26.7% |
| 5000 | 20 | 0.4% | 0.8% |
|  | 50 | 1.0% | 2.0% |
|  | 100 | 2.0% | 4.0% |
|  | 200 | 4.0% | 8.0% |
| 8192 | 20 | 0.2% | 0.5% |
|  | 50 | 0.6% | 1.2% |
|  | 100 | 1.2% | 2.4% |
|  | 200 | 2.4% | 4.8% |

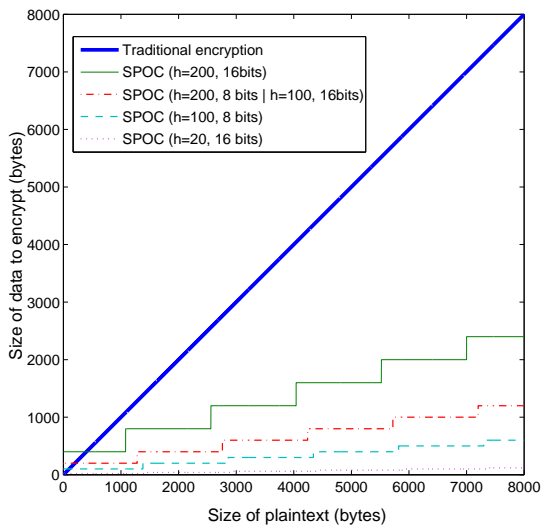

Fig. 4. Size of data to be encrypted, for SPOC (encryption of locked coefficients) versus traditional encryption (encryption of the whole data).

## IV. PERFORMANCE EVALUATION

Seeking performance evaluation criteria that are technology independent, we focus on encryption volume, space requirements and computational overhead of SPOC.

*Encryption Volume:* To illustrate SPOC's efficiency in comparison to traditional end-to-end encryption, *Figure 4* compares the volume of data to be encrypted according to the size of the plaintext. We consider a maximum payload size of 1480 bytes (a typical value e.g. for the Ethernet) and assume that each generation encoded by SPOC has $h$ coded packets. In the case of the traditional encryption mechanism, which performs end-to-end encryption of the entire payload, the volume of data that must be encrypted increases linearly with the size of the protected payload. It is not difficult to see that, by encrypting solely the locked coefficients, SPOC substantially reduces the size of information to be encrypted (both in the case of 8 bit and 16 bit coefficients), while still guaranteeing strong confidentiality of the payload.

Notice that the size of information to be encrypted by our scheme increases in discrete steps, because it is only when the payload size surpasses the maximum payload size of a typical IP packet that a new set of $h$ locked coefficients must be generated and included in the next IP packet. It is also important to notice that if we consider IP packets which can contain more than the maximum payload size of typical IP packets (1480 bytes), the relative gains of our scheme are deemed even higher than for this specific case. This happens because more data can be sent in each packet containing the same set locked coefficients.

Naturally, the required number of cryptographic operations is directly related to the aforementioned volume of data (to be encrypted). If we consider a stream cipher, the number of operations to perform while encrypting increases linearly with that volume, and therefore, the number of operations is greatly reduced by SPOC as shown in *Figure 4*. In the case of a block cipher such as AES (with blocks of length 128 bits) each encryption operation allows up to 128 bits of input data, i.e. 16 coefficients in a field of size $2^8$, or 8 coefficients in a field of size $2^{16}$.

*Space requirements:* The ability to reduce the volume of data to be encrypted comes at the cost of including locked coefficients in the data packet. Assuming the use of an encryption mechanism in which the ciphertext has the same size of the plaintext (e.g. AES in stream cipher mode), the overall transmission overhead for a generation of $h$ coded packets depends on the maximum IP packet size.

In *Table II* we show the overhead introduced by SPOC for each packet and for coefficients with size of 8 and 16 bits. The overhead values are obtained by calculating the ratio between the size of the locked coefficients and the maximum size of an IP packet. As the number $h$ of packets to encode increases, so does the packet size overhead, since each new packet to encode adds a new locked coefficient to the IP packet. For high values of $h$, the overhead can become significant, in particular for IP packets of small size (e.g. coding 200 packets and taking a maximum IP packet size of 1500 bytes (including the 20 byte header) results in a overhead of 13.3% in a field with size $2^8$ and 26.7% in a field with size $2^{16}$). However, this overhead can be significantly reduced by increasing the packet size (e.g. for the same 200 coded packets, the overhead can be reduced until 2.4% in a field with size $2^8$ and 4.8% in a field with size $2^{16}$).

TABLE III
COMPUTATIONAL COST OF INCLUDING THE LOCKED COEFFICIENTS, PER GENERATION OF SIZE $h$.

| NODE | OPERATION | DETAILED COST | TOTAL COST |
|---|---|---|---|
| Source Node | Generation of vectors of identity matrix | negligible | $O(h^2)$ |
| | Encryption of locked coefficients | See *Section IV* | |
| | Performing extra random linear operations | $h^2$ multiplication and $(h-1)h$ sums | |
| Intermediate Node | Performing extra random linear operations (combining $n$ packets) | $nh$ multiplication operations and $(n-1)h$ sum operations | $O(nh)$ |
| Sink node | Inverse of the unlocked coefficients' matrix $M_U$, $M_U^{-1}$ | $O(h^3)$ | $O(h^3)$ |
| | Product $M_P$ of $M_U^{-1}$ with the locked coefficients' matrix $\hat{M}_L$, $M_P = M_U^{-1} \times \hat{M}_L$ | $O(h^3)$ | |
| | Decrypt locked coefficients to obtain the matrix $M_L$ of plain-text locked coefficients | See *Section IV* | |
| | The product $M = M_U \times M_L$, to obtain the final matrix in which Gaussian Elimination will be performed | $O(h^3)$ | |

*Computational overhead:* Due to the inclusion of an extra set of coefficients (the locked coefficients), SPOC introduces additional operations at network nodes that do not exist in standard RLNC based protocols. The computational overhead is different for the source, intermediate and destination nodes. In *Table III*, we provide a detailed evaluation of the computational overhead of SPOC.

For the purpose of our analysis, we consider that, in comparison to the multiplication, the sum operation yields negligible complexity. We consider a naïve approach for matrix multiplication, which takes $O(n^3)$ operations. Although there exist algorithms with lower computational complexity, the resulting improvements are minor, and usually apply only to very large matrices. Since including large sets of packets in each generation (i.e. large $h$) requires us to perform Gaussian elimination on large matrices and increases the overhead of network coding protocols (in terms of additional coefficients to include in a packet), we consider this approach for matrix multiplication reasonable for the scope of our analysis.

## V. FINAL REMARKS

We presented a security scheme that assures confidentiality in network coding protocols based on the interplay between the coding properties in this paradigm and cryptographic mechanisms. Specifically, we attained a substantial reduction on the size of the data to be encrypted when compared to the naïve encryption approach (where the whole data needs to be encrypted) and, consequently, a reduction of the computational overhead required to perform encryption. Confidentiality is achieved by protecting (or "locking") the source coefficients required to decode the linearly coded data, and by letting intermediate nodes run their operations on a set of coefficients composed by the "unlocked" and the "locked" coefficients that do not impair any of the operations of practical network coding protocols. Active attacks can be countered by the use of a shared secret between source and sinks, to help detect the errors introduced by the attacker with inherent capacity and complexity trade-offs. As part of our ongoing work, we are considering the interplay between our network coding security scheme and priority encoding transmission, as well as security mechanisms to address Byzantine attacks on network coding.

## REFERENCES

[1] R. Ahlswede, N. Cai, S. Li, and R. Yeung, "Network information flow," *IEEE Transactions on Information Theory*, vol. 46, no. 4, pp. 1204–1216, 2000.
[2] R. Koetter and M. Médard, "An algebraic approach to network coding," *IEEE/ACM Trans. Netw.*, vol. 11, no. 5, pp. 782–795, 2003.
[3] D. Lun, M. Médard, R. Koetter, and M. Effros, "Further results on coding for reliable communication over packet networks," *IEEE International Symposium on Information Theory (ISIT), Adelaide, Australia*, pp. 1848–1852, September 2005.
[4] T. Ho, M. Médard, R. Koetter, D. Karger, M. Effros, J. Shi, and B. Leong, "A random linear network coding approach to multicast," *IEEE Transactions on Information Theory*, vol. 52, no. 10, pp. 4413–4430, 2006.
[5] S.-Y. R. Li, R. W. Yeung, and N. Cai, "Linear network coding," *IEEE Transactions on Information Theory*, vol. 49, no. 2, pp. 371–381, 2003.
[6] P. Chou, Y. Wu, and K. Jain, "Practical network coding," *Allerton Conference on Communication, Control, and Computing, Allerton, USA*, September 2003.
[7] C. Gkantsidis and P. Rodriguez, "Network coding for large scale content distribution," *Proceedings of IEEE Infocom, Miami, USA*, 2005.
[8] A. G. Dimakis, P. B. Godfrey, M. J. Wainwright, and K. Ramchandran, "Network coding for distributed storage in peer-to-peer systems," *In Proc. of IEEE INFOCOM, Anchorage, Alaska*, May 2007.
[9] J. Widmer and J.-Y. L. Boudec, "Network coding for efficient communication in extreme networks," in *WDTN '05: Proceeding of the 2005 ACM SIGCOMM workshop on Delay-tolerant networking*. New York, NY, USA: ACM Press, 2005, pp. 284–291.
[10] S. Katti, H. Rahul, W. Hu, D. Katabi, M. Médard, and J. Crowcroft, "XORs in the air: practical wireless network coding," in *SIGCOMM '06: Proceedings of the 2006 conference on Applications, technologies, architectures, and protocols for computer communications*. New York, NY, USA: ACM Press, 2006, pp. 243–254.
[11] N. Cai and R. Yeung, "Secure network coding," *IEEE International Symposium on Information Theory, Lausanne, Switzerland*, July 2002.
[12] J. Tan and M. Médard, "Secure Network Coding with a Cost Criterion," *Proc. 4th International Symposium on Modeling and Optimization in Mobile, Ad Hoc and Wireless Networks (WiOpt'06), Boston, Massachussets*, April 2006.
[13] L. Lima, M. Médard, and J. Barros, "Random linear network coding: A free cypher?" in *In Proc. of the IEEE International Symposium on Information Theory, Nice, France*, June 2007.
[14] S. Jaggi, M. Langberg, S. Katti, T. Ho, D. Katabi, and M. Médard, "Resilient Network Coding In the Presence of Byzantine Adversaries," *In Proc. of IEEE INFOCOM 2007, Anchorage, Alaska*, May 2007.
[15] C. Gkantsidis and P. Rodriguez, "Cooperative security for network coding file distribution," *In Proc. of IEEE INFOCOM, Barcelona*, April 2006.
[16] R. Koetter and F. Kschischang, "Coding for Errors and Erasures in Random Network Coding," *IEEE International Symposium on Information Theory (ISIT), Nice, France*, June 2007.